# Social Media and the Social Good: How Nonprofits Use Facebook to Communicate with the Public

# 社交媒体与社会公益：非营利组织如何使用"脸书"与公众沟通

Gregory D. Saxton; Chao Guo; I-Hsuan Chiu; Bo Feng

# 社交媒体与社会公益：非营利组织如何使用"脸书"与公众沟通？[*]

Gregory D. Saxton博士

美国纽约州立大学水牛城分校传播学系

郭超博士

美国乔治亚大学公共行政与政策系

邱宜轩

美国纽约州立大学水牛城分校传播学系

冯博

美国乔治亚大学公共行政与政策系

**摘要**

本文以"脸书"为例，研究美国 100 家最大的非营利组织如何利用社交媒体来改善与公众之间的沟通互动。我们发现，非营利组织在其"脸书"主页上的"状态"更新可以被归纳为三种核心功能："信息传播功能"，"社区建设功能"和"动员促进功能"。我们的分析表明，尽管"脸书"的"信息传播功能"使用程度依然相当显著，但比起传统意义上的网站，非营利组织能够通过"脸书"中的"社区建设"和"双向对话"功能更有效地与公众沟通。社交网络的应用似乎树立了组织与大众互动的新范式。

**关键词**：社交媒体；互联网；非营利组织；"脸书"；公共关系；组织沟通。

互联网的快速普及为非营利组织与公众之间的沟通提供了新的可能性。然而，近来的研究（e.g., Kent, Taylor, & White, 2003; Saxton, Guo, & Brown, 2007）却表明非营利组织未能充分利用组织网站来更有效地实现与公众之间的互动与交流。造成这种不足的原因可能是由于非营利组织缺乏与互联网有关的专业技术和专业人才，从而没有能力建立起反馈信息和参与讨论的网络互动平台。但随着"脸书"（Facebook）和"推特"（Twitter）等各种免费社交媒体的兴起，任何组织，无论大小，都可以在社交媒体上面建立一个主页，并通过主页与公众进行即时、有效的沟通。

本研究中，我们将针对非营利组织如何使用社交媒体进行公众沟通并鼓励公众参与这一问题进行调查。具体而言，我们通过对美国100家最大的非营利组织在其"脸书"主页上的"状态"更新的频率与内容的分析，来了解社交媒体在非营利组织中所承担的功能。我们进一步将状态更新根据其内容归纳为三个核心功能：信息传播功能，社区建设功能，和动员促进功能。与以往对网站的研究发现相反，我们的研究表明，非营利组织已经开始利用社交网络来实现一种具备多重"对话"和"互动"功能的公众沟通方式。

本文共分四部分。第一部分回顾现存文献，探讨社交媒体在非营利组织公共关系管理方面的潜能；第二部分介绍研究方法与数据；第三部分对研究结果进行分析；第四部分讨论本文的贡献与局限以及未来研究的方向。

## 社交媒体作为非营利组织与公众沟通的工具

互联网为非营利组织传递信息以及与公众相互沟通提供了崭新而宝贵的资源。互联网直接、高效和经济的特点对于那些资源相对贫乏的非营利组织显得尤为重要

（Young and Salamon, 2002; Waters, 2007）。另外，互联网 2.0 技术所带来的强大的互动性和参与性为非营利组织提供了一种新型的组织沟通战略，使非营利组织可以更好地与资助者、客户、媒体以及公众相互交流（e.g., Hackler & Saxton, 2007; Kang & Norton, 2004; Saxton, Guo, & Brown, 2007）。

近来，越来越多的研究开始致力于探讨以下问题：非营利组织在多大程度上能够利用互联网来改善与公众之间的沟通互动，从而实现组织的使命与目标？具体而言，信息和组织运营的公开化以及资金筹集和双向对话的能力已经被公认为非营利组织公众关系管理的核心因素（e.g., Saxton, Guo, & Brown, 2007; Taylor, Kent, White, 2001; Waters, 2007）。有证据显示，提供组织信息以及公开组织运行状况（尤其是财政状况）是与公众搭建信任桥梁的两个最基本方式（Saxton & Guo, 2011; Waters, 2007）。随着大家对互联网的依赖与日俱增，在决定对某个非营利组织捐赠善款或者担任志愿者的时候，人们越来越借助网络来了解相关的组织信息（Gordon, Knock, & Neeley, 2009; Spencer, 2002）。

对话式互动或者双向沟通被视为互联网能提供的最诱人的功能，也被认为是促进非营利组织与公众相互合作的重要工具（Saxton, Guo, & Brown, 2007）。近期的研究也着手开始探讨非营利组织如何利用互联网 2.0 技术的互动功能来改善与利益相关者（stakeholders）之间的关系（e.g., Ingenhoff & Koelling, 2009; Kang & Norton, 2004），以及提高组织在资金募集（e.g., Sargeant, Douglas, West, & Jay, 2007; Yeon, Choi, & Kiousis, 2005）、慎议式公共空间（e.g., Kenix, 2007）以及虚拟社会资本（e.g., Nah, 2009）等方面的能力。

以互联网 2.0 技术为基础，博客（blogs）、"我的空间（MySpace）"、"推特（Twitter）"以及"脸书（Facebook）"等一系列的社交媒体已经为改善组织同公众之间的沟通与互动关系创建了一个新的平台。社交网络以分享、合作和动员公众的方式同公众构建合作关系（Greenberg & MacAulay, 2009）。社交媒体为公众参与讨论提供了一个低成本的途径，从而将那些可能被传统媒体忽略的问题重新拉入公众眼帘。这样一种在线式的互动环境被认为是构建个人与组织双向对话的一个合适选择（Bortree & Seltzer, 2009; Lovejoy, Waters, & Saxton, forthcoming）。

目前，对社交媒体在非营利组织中运用的研究仍然非常有限。Seltzer & Mitrook（2007）用两个不同的数据来对比 50 个博客账户和 150 家环境保护组织的网站，发现人们更多是选择博客而并非传统网站作为彼此间相互沟通、联系的方式。研究者还进一步发现，比起传统网站，博客具有更多的对话功能，因而是一个更好的双向交流工具。Lovejoy & Saxton（2011）的研究发现，美国 100 家最大的非营利组织已经开始利用"推特"来发展双向对话和社区建设。另外，Briones, Kuch, Liu, & Jin（2011）访谈了 40 位红十字会的工作人员，也发现"推特"等社交媒体在培养和加强组织与公众之间的关系方面存在重要的潜能。

对"脸书"在非营利组织间使用情况的研究更是寥寥无几。Bortree & Seltzer（2009）对 50 个环境保护组织的"脸书"账户进行调查后，发现很多组织错误地以为在社交网站上建立主页就等于改善了组织与公众之间的沟通与互动。Waters, Burnett, Lamm & Lucas（2009）对 275 家非营利组织的"脸书"主页进行了调查，发现这些组织在"脸书"主页上的信息大都局限于描述组织本身的情况，而往往忽略了

关系构建的功能。Greenberg & MacAulay（2009）对 43 家加拿大环境保护组织在传统网站和"脸书"，"推特"以及"博客"等其他社交网络中的使用情况做了调查。他们发现有过半数的组织拥有"博客"账户，然而很少的组织拥有像"脸书"或"推特"这样的社交媒体的账户。尽管越来越多的非营利组织开始使用社交媒体来改善他们沟通和资金募集的战略方式，但是以上的三个研究报告却说明这些组织并未能充分利用社交媒体这一功能。

总的来说，社交媒体以其区别传统网站的高度交互式功能而成为非营利组织改善公众关系的一种潜在的重要手段。然而，当前有关非营利组织在线沟通和互动战略研究的文献往往只关注它们对组织官方网站的使用而忽视了组织对社交媒体的使用。因此，我们需要进一步了解社交媒体的独特功能，并且研究这些功能如何加速和促进组织与公众之间的互动。

本文旨在提供一种新的框架来评估非营利组织与公众之间构建互动关系的策略。具体而言，我们通过研究非营利组织在"脸书"主页上的"状态更新"（status updates）来探讨它们是如何与公众之间建立一个公开、透明、互动的关系。我们尝试对下述问题作出解答：非营利组织会通过社交媒体发布哪些种类的信息？组织对社交媒体所提供的各种功能的依赖程度如何？

## 方法论

### 抽样

考虑到社交媒体是一种新型的组织沟通工具，我们选择对大型慈善机构进行抽样。尽管对"脸书"的使用并不要求雄厚的经济实力和强大技术支持，但比起其他

小型组织，大型机构对"脸书"的使用显得更加普遍。我们的样本来源是美国 2008 年"非营利组织百强"。"非营利组织百强"的名单每年发表一次，依据年度财政收入评选出美国前 100 家最大的非教育类非营利组织。一家非营利组织要入选百强，其年度财政收入至少有 10%需要来自于公众的私人捐助。

**过程**

我们为本研究专门编写了 Python 程序用来下载每一个组织在"脸书"主页上的"状态"更新。为了检验 Python 下载数据的准确性，我们做了一个试验下载并将 100 家组织随机性地与彼此在"脸书"上的状态数据进行对比。在确保数据的准确性之后，我们于 2009 年 12 月 5 日至 2010 年 1 月 4 日期间下载了所有组织在"脸书"主页上的状态更新数据。虽然总共抓取到的状态更新数据有 1,043 个，但其中有 7 个冗余数据。在剔除冗余数据后，我们的总数据为 1,036 个。

为了对组织状态更新的数据进行内容分析（content analysis），我们依据相关文献以及状态更新的内容设计了一套编码方案；在这一编码方案里，我们关注的是每一次状态更新所承担的核心功能。我们首先各自独立地对 54 个状态更新数据进行编码，然后在对比和讨论这 54 个状态更新数据的基础上，我们对编码做了进一步的修改。此后，我们开始对另外 103 个状态更新数据进行编码，并以此用作内部数据可靠性的对比。对比结果显示出高度的可靠性（93.2% inter-coder agreement；a Cohen's kappa score of .89）。最后，我们对全部 1,036 个数据进行了编码。

**结果**

在"脸书"为用户提供的诸多功能之中，其核心特色与两种功能有关："涂鸦墙"（The Wall）和"状态"（Status）。"涂鸦墙"像是一个互动的留言板：在这个留言板上，用户和"粉丝"（fans）可以公开地彼此交换信息。而"状态"则是让用户在"涂鸦墙"上发布信息，向"粉丝"和Facebook社区显示自己在哪里、做什么。状态的一般格式包括发帖者本人的照片以及发帖者所传递的内容，并且这些信息对所有注册用户都是公开的。这样，非营利组织可以通过及时更新自己的"状态"来向公众和"粉丝"发布组织的最新动态。在某种程度上，"脸书"的"涂鸦墙"和"状态"功能可以很好地把组织与"粉丝"有效地联系起来，围绕组织的使命和目标形成一个虚拟社区。在这里，我们通过对组织官方"脸书"主页上状态更新的频率与内容的分析，来了解社交媒体在非营利组织中所承担的功能。

我们发现在100所组织中，有65所拥有一个固定的官方"脸书"主页。在数据收集期间，每家组织发布状态更新的平均频率为10.44次。图1显示的是65个组织所发布状态更新的频率。可以看出，组织间在状态更新的频率上存在很大差别。

------------------------------

图1

------------------------------

"脸书"的"涂鸦墙"功能同样也使得"粉丝"与组织间的交流变得透明起来，因为我们可以看到"粉丝"对组织状态的评价是"鲜花"还是"鸡蛋"。与此同时，对"粉丝"提出的问题或是要求，组织也可以通过状态更新进行回应。

**状态的分类**

基于对状态更新这一核心功能的分析，我们界定了五种基本类型的状态更新："信息传播"，"筹集善款"，"事件宣传"，"动员促进"，和"社区建设"。表1指明每一种状态的出现频率。对那些不属于以上五种基本类型的状态，我们把他们统一归为"杂项"这一类。因为这一类的出现次数较少，我们也就不再单独进行讨论。通过归纳分析，我们进一步将这五类状态更新类型合并为三个主要的维度："信息传播"，"社区建设"和"动员促进"。这三个维度反映了状态更新的三种基本功能。第一个功能："信息传播功能"，是指组织动员通过"脸书"主页的状态更新来传播与组织本身及其相关活动有关的信息。第二个功能："社区建设功能"，是指组织通过状态更新来培养与公众的关系，以及共建社区。第三个功能，"动员促进功能"，是指组织通过状态更新来动员公众为组织采取各种行动，诸如募捐，参加活动，甚至是发起一场抗议。下面，我们对这三个维度的状态更新做进一步的讨论。

-------------------------------

表 1

-------------------------------

**信息传播功能**

"信息传播"类的状态更新占据了全部状态更新的 51.68%。这一类型的状态更新以传递与非营利组织相关的新闻和组织本身的信息为主，包括：组织本身的使命，其实时更新的活动内容，其工程开展的进度，以及最近的媒体曝光。该类型的更新往往会通过链接的方式提供更加具体、更加准确的信息。与先前对大多数网站的调查结果相似的是，这一维度的状态更新展示了非营利组织在传播信息上的强大实力（e.g., Kenix, 2007; Kent et al., 2003; Saxton et al., 2007; Waters, 2007）。通过公开与

自身职责，背景，活动及绩效相关的信息，组织可以培养其公众责任感（e.g., Spencer, 2002; Waters, 2009），也可以更好的培养与公众及公众之间的互信关系。正如 Spencer 在 2002 强调的，责任感是非营利组织与公众和公众培养战略关系的总要工具。信息传播类的状态更新的例子包括：

> 请用 2 分钟的时间观看我们为过去 5 年里海啸救援所制作的一组幻灯片：
> 
> http://crs.org/indonesia/rebuilding-communities/Aceh, Indonesia, Five Years Latercrs.org。2004 年 12 月 26 日的海啸不仅首先袭击了印尼亚齐省，而且给亚齐省带来了巨大的损失。天主教救急服务组织于事发当天立即奔赴救灾现场，为将近 25 万难民发放食物，帐篷，医疗卫生工具等一些生活必需品……
> 
> 我们在过去两天里收到了两份有意思的年终捐赠。（第一份捐赠是，）奥巴马总统和第一夫人继续对我们的工作给予支持。第二份同等重要和特别的捐赠来自圣何塞的一位名叫伊兹 厄尔曼的 13 岁小女孩，她寄来了一张通过经营宠物关怀而赚得的 1000 美金的支票。

**社区建设功能**

这一类型的状态更新占全部状态更新的 15.72%，主要包括：故事分享，人物致谢，提出请求，问卷调查，小测验，以及节日祝贺，等等。例如，故事分享类的状态更新，以描述贫困人民的故事或分享志愿者的经历为依托，旨在提高公众对组织活动的关注度与参与度。这一环节通过与公众构建双向对话，搭建关系桥梁，最终建立一个网络社区。鉴于非营利组织在互联网时代里所长期表现出的与公众缺乏双向互动的窘境（e.g., Kent et al., 2003; Ingenhoff & Koelling, 2009; Kang & Norton, 2004），社区建设功能似乎可以通过双向交流来改善当前的局面。下面请看两个例子：

我们发明了一个新的测试！你曾经幻想过自己是哪一种动物吗？请点击加入我们的测试：http://bit.ly/AnimalQuiz

新年快乐！我们再次回到工作岗位，蓄势待发。2010年伊始，你希望从我们的"脸书"主页上看到什么样的内容？是更多的新闻？图片？留言？讨论？还是其他更多？请速与我们联系。

## 动员促进功能

这一类型的状态更新旨在动员公众和"粉丝"为组织采取行动。这样一种贡献既可以是参与其中，提出倡导，也可以是资金募捐。它将已行动为向导，整合各种在线或本地资源，来帮助组织实现其经济和战略目标。总体而言，在我们的样本中已有32.22%的状态更新实现了动员促进这一目标。

*募集善款*。这些状态旨在从公众中筹集到更多的捐款。募捐的途径很多，既可以直接募捐，或者购买拥有组织版权的商品，也可以以参与游戏的方式来赢得奖金，从而再次募捐给组织。可以说，参与游戏的方式并非一个直接的经济资助，相反，它是通过公众的参与从侧面来为组织赢得善款。这样的一种方式，为那些在经济上无法直接捐助组织的人们提供了另外的一种途径。例如：

> 尽管你没有实物可以直接捐助给当地的Goodwill，但你想在帮助自己免税的同时也帮助他人克服就业难题吗？那么请在线资助我们来帮你完成这一使命吧。请通过Onlineeservices.goodwill.org来捐款给我们。您对Goodwill的经济支助将帮助我们为更多的人实现就业梦想，帮助他们改善生活，帮助他们建设自己的家庭和社区。

*活动宣传*。这些状态用来宣传组织近期的活动以及鼓励公众通过各种方式参与其中（例如身临现场，收听广播，观看电视，或在线交流）。此外，有些组织还可以通过发放免费电子卡片（有别于以募捐为目的的电子卡片）给支持者，并通过他们的社交圈子传播给更多的人。例如：

> 请不要错过我们于 12 月 24 日下午 5 时 30 分在位于 Historic Area 的 Market Square 举行的平安夜圣诞树结彩仪式。传统的圣诞颂歌和一些特别故事将贯穿今年的结彩仪式……本次活动可以免费入场。点击观看更多……

*号召行动*。这一类的状态更新仅仅构成我们样本中的一小部分，约占全部状态更新的 2.88%。这一类状态更新的目标很明确，就是号召公众与"粉丝"参加游说、倡导以及其他志愿行动。此外，此类状态更新也鼓励组织的支持者向亲朋好友来宣传组织本身，从而扩大组织支持者的数量。鼓励支持者为组织宣传造势，鼓励支持者去邀请他们的朋友成为这个组织新的支持者，或者让支持者使用这个组织的图标作为自己"脸书"的图像。种种行为都有助于组织的壮大。例如：

> 动员你们的参议员，让他支持我们的 Wyden-Durbin 提议案。从而让那些病人在不损失公共医疗福利的情况下也能免费接受诊所的治疗。你可以通过如下方式联系到你身边的参议员：http://bit.ly/8WhIqG

总体而言，以上的三个维度涵盖了组织"脸书"主页的状态更新所承载的主要功能，也恰好对应了以往在非营利组织构建关系战略方面的研究发现——提供信息，募集资金和双向对话。尽管有半数的状态更新是属于纯粹的信息传播，但是另一半的状态更新却包括了动员和对话。因此，"脸书"拓宽了组织与公众互动合作的渠道。

**组织的分类**：

在上一个部分，我们将组织"脸书"主页的状态更新进行了分类。现在我们转向组织层面的分析，来看一看是否可以依据状态更新的特点来区分不同类型的组织。例如，根据组织对"脸书"主页状态更新的主要功能的依赖程度不同，我们可以发现在"脸书"中是否存在真正的"对话"型组织。

为了实现这一目标，我们对组织对于状态更新的三个主要功能——信息传播、社区建设和动员促进——的相对依赖程度进行了分析。从下面一个三角图可以清晰地看出不同组织对于上述三个功能的相对依赖程度的差别（参看图 2）。

------------------------------

图 2

------------------------------

三角图中有 49 个点；每一个点代表一个在采样阶段比较"活跃"的组织。我们对"活跃"的定义是：在采样阶段，每周至少在组织的"脸书"主页进行三次状态更新。在这个三角图中，不同位置的点代表该组织的状态更新在信息传播，社区建设，和动员促进等功能中各自所占的不同比重。一个位于三角图顶端的点将表示这个组织的状态更新全部是信息传播，从而那些离顶端越远的点将代表这个组织越少的状态更新是信息传播。当一个点位于三角图最底端时，它也就意味着这个组织所有的状态更新都是非信息传播类的。与之类似，一个位于右下角的点意味着这个组织所有的状态更新都是社区建设，而位于左下角的点意味着这个组织所有的状态更新都是动员促进。那些位于三角图中心的点代表着三个功能等分的情况。

通过对三角图的分析，我们可以有效地识别出三种功能在每一个组织中各自的份额，进而帮助理解非营利组织在管理"脸书"主页方面的行为类型。我们的做法与 Java et al. (2007) 在他们研究中对"推特"用户进行分类的方法类似。基于用户发布"推特"消息的不同习惯和动机，作者将用户分为以下三类："信息提供者"，"朋友"和"信息采纳者"。同理，我们将"脸书"的非营利组织用户分为三类："信息提供者"，

"宣传动员者"，和"社区建设者"。汇聚于三角图中心的三条线将三角图等分为三个部分，从而区分出三种不同类型的组织。

因为大多数的点都汇聚在"信息传播"部分，可见大多数的组织（31个）属于"信息提供者"。这也证明组织中有很大程度的状态更新是信息传播的更新。在"宣传动员者"部分有13个组织，而在"社区建设者"部分则一个组织也没有。在三角图中，有5个组织坐落于"信息"和"行动"之间，这也就意味着这些组织在"信息传播"和"动员促进"方面的状态更新是等量的。特别来说，图中左侧的6个组织标志着他们的状态更新兼具"信息传播"和"行动促进"，但缺乏"社区建设"的特点。其他3个位于右侧的组织也就意味着他们的状态更新兼具"信息传播"和"社区建设"，但缺乏"动员促进"的特点。

在很大程度上，样本中的非营利组织大都将"脸书"主页的重心放在为公众提供信息上，其次是动员公众和"粉丝"为组织采取行动，再次是构建一个共同的社区。这一发现似乎证实了以往的研究，即大多数非营利组织尚未成分发掘利用社交媒体的对话与互动的功能（e.g., Bortree & Seltzer, 2009; Waters et al., 2009）。然而，进一步的分析表明非营利组织开始在逐步地利用社交媒体的双向沟通这一功能。

## 结论

本文以"脸书"为例，研究美国100家最大的非营利组织如何利用社交媒体来改善与公众之间的沟通互动，达成组织的使命与目标。通过对组织在其"脸书"主页上状态更新的分类，我们界定了三种主要的公共关系建设战略："信息传播"，"动员促进"和"社区建设"。信息传播仍然是使用频度最高的策略，但是正如我们在诸如呼

吁捐赠与志愿活动以及鸣谢与鼓励公众献计献策等等状态更新中所看到，宣传倡导与双向互动已经成为一种趋势。

我们的研究发现，非营利组织已经开始利用社交媒体这个新平台来改变他们交流、协调、宣传倡导和资金募集的方式。尽管美国前 100 家非营利组织中只有三分之二的组织拥有"脸书"主页，但是非营利部门为运用高新科技而作出的努力已经有所展现。有限的资源和组织自身的文化也许可以解释为什么他们至今还未充分利用社交媒体：尽管社交网络成本很低，但对它的维护和升级确是一个繁琐的过程。在某些情况下，还有可能导致过高的管理经费，从而更多的资源需要被分配。此外，正如 Greenberg 和 MacAulay 在 2009 年所指出的，并非每一个非营利组织都必须向他们的公众或者向大众做出一种双边沟通的承诺。相反，他们也许将资金募集，倡导或者会员管理放在首位以实现他们的战略目标。

非营利组织可以运用社交媒体与公众进行更加有效的沟通与互动。随着社交媒体与我们日常生活的联系日益紧密，一个跨年龄、跨文化、跨地域的多元化群体正在被带入我们的生活圈（Waters et al., 2009）。与此相适应，非营利组织也应当更充分地利用社交媒体来联系公众以及服务公众需求。如先前研究所指出，组织自身的目标与实现网络关系的构建以及实际生活中的双边互动这三者间还是未能同步（e.g., Bortree & Seltzer, 2009）。因此，这就要求非营利组织必须付出更多努力（包括时间、财力和人力方面的重新分配）来改进这一现状。

需要指出的是，我们的研究存在一些重要的局限。首先，因为我们的研究仅涉及对 65 个"脸书"账户的调查，样本很小。其次，样本的元素均来自于美国前 100

家大型的非营利组织，他们并没有包括其他小型或者以社区为单位的非营利组织。再者，非营利组织与用户间的"悄悄话"（private messages）无法被我们获取，因此它们都未能纳入我们的分析。尽管，一对一的"悄悄话"是组织与公众一种的好方法，但是如果"悄悄话"是用户更为偏好的沟通方式，那么这可能会部分影响到组织"脸书"主页上的能够观测到的公众沟通与互动的比重。如此，作出非营利组织尚未有效利用"脸书"与公众进行有效的沟通的结论可能过早。

除了考虑上述的局限之外，今后的研究应当进一步探讨不同的组织对于不同公共关系建设——传递信息，宣传/动员和社区建设——的依赖程度。例如，我们可以做一个调查：是否倡导性组织或者志愿者组织会更加青睐于发布动员促进类的消息，而并非传统的信息传播？另外，还可以研究非营利组织如何使用除"脸书"以外的其他社交媒体来实现组织的战略目标。最后需要指出的是，交流是一个双向过程，而本文仅仅对非营利组织方面的行为进行了调研。因此，未来的社交媒体研究可以考虑公众如何看待和回应组织在公共关系建设方面的努力。

"他山之石，可以攻玉。"研究西方国家非营利组织对社交媒体的运用对于中国非营利部门也具有启迪与借鉴意义。目前，中国非营利组织仍处于资源匮乏的起步阶段，而社交媒体或许可以成为非营利部门发展壮大的新的推力。作为当今全球最大的社交媒体，"脸书"的影响力自不待言。而"人人网"（前身为"校内网"）——中国本土开发的"脸书"——同样给无数的个人，组织和机构带来意料之外的便利和经济效益。非营利组织应当充分利用"人人网"强大的信息传播与互动功能，与公众建立起愈发健康、和谐的双向沟通关系。至于如何运用社交媒体来实现自身的战略目

标，途径和方法有很多。比如，非营利组织可以在"人人网"上注册自己的公共主页，并通过公共主页上类似于"脸书"的状态栏发布其最新活动消息；组织可以通过发布日志，通过现存关注用户转发日志让更多的"人人网"用户看到组织的活动信息，从而扩大组织的"粉丝"群；而通过"人人网"进行人才招聘，也是扩大组织影响力的一个有力措施。

参考文献

表 1 状态更新的功能分类

| 分类 (Category) | 频率 (Frequency) | 比例 (Percent) |
| --- | --- | --- |
| **信息传播类 (Informational )** | | |
| 信息传播 | 536 | 51.74 |
| **动员促进类 (Promotional/Mobilizational)** | | |
| 筹集善款 | 209 | 20.17 |
| 活动宣传 | 95 | 9.17 |
| 号召行动 | 29 | 2.80 |
| **社区建设类 (Dialogic & Community-Building)** | | |
| 社区建设 | 163 | 15.73 |
| **杂项类 (Miscellaneous)** | | |
| 杂项 | 4 | 0.39 |
| 总计 (Total) | 1,036 | 100% |

**图 1：组织状态更新的频率, 2009 年 12 月 5 日 – 2010 年 1 月 4 日**

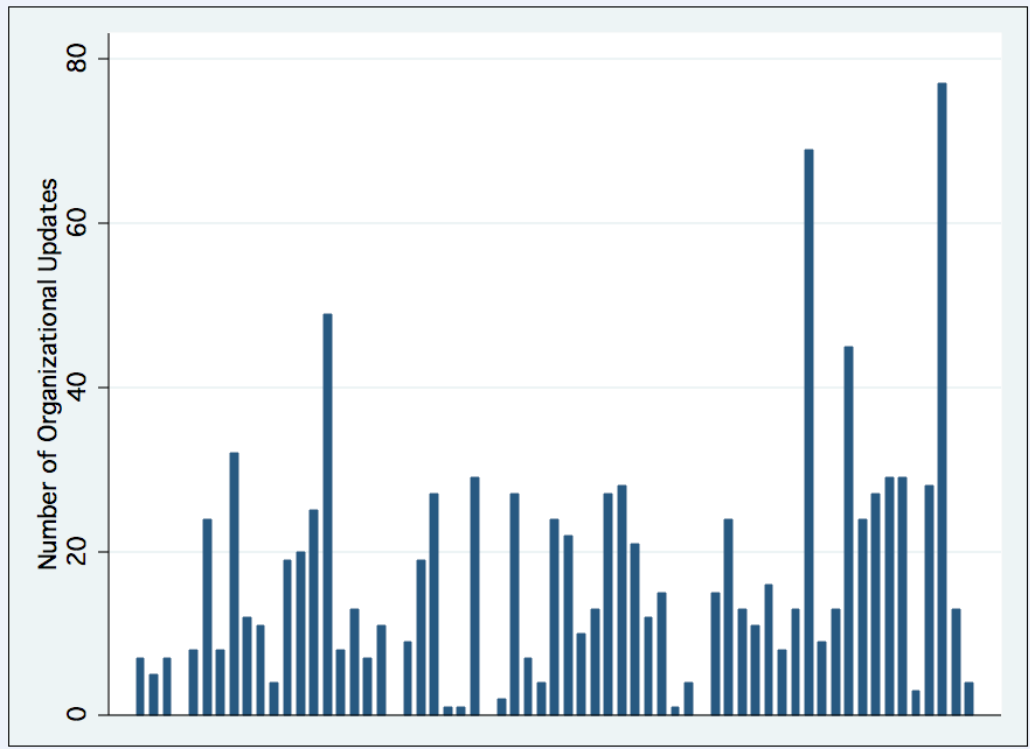

**图 2： 三角图: 组织对状态更新的三个核心功能的依赖程度**

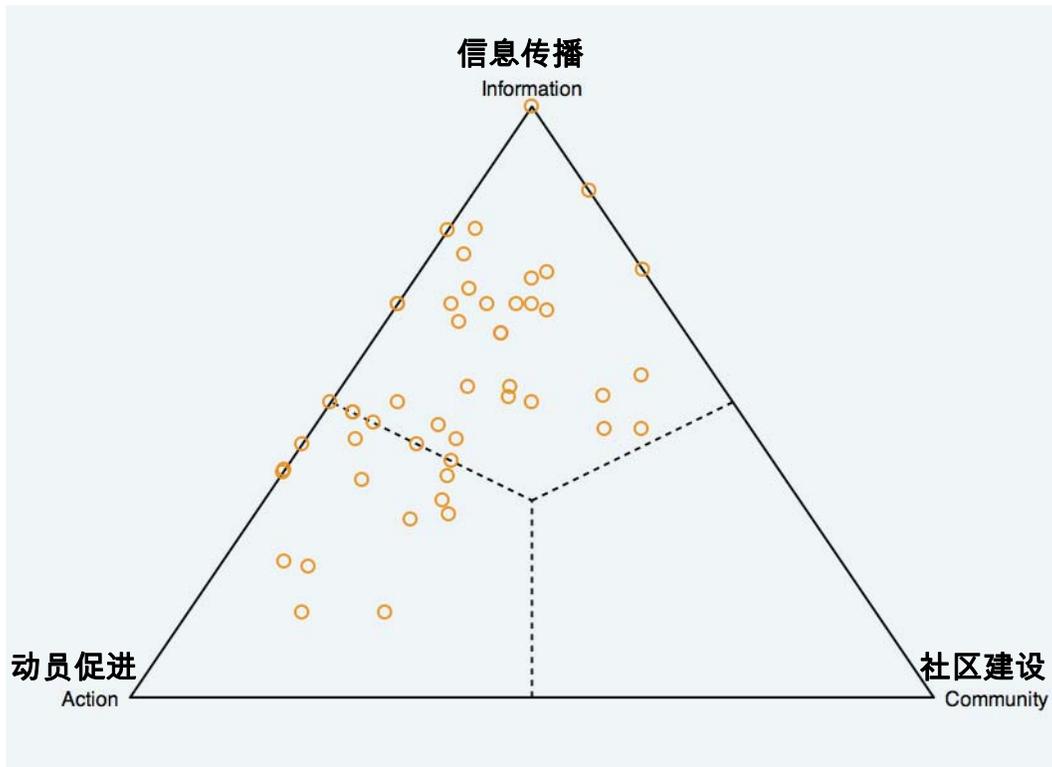